\documentclass{bmcart}
\pdfoutput=1

\usepackage{amsthm,amsmath}

\usepackage{graphicx}

\startlocaldefs

\usepackage{glossaries}
\usepackage{amssymb}
\usepackage[utf8x]{inputenc}

\newacronym{osa}{OSA}{opportunistic spectrum access}
\newacronym{pu}{PU}{primary user}
\newacronym{su}{SU}{secondary user}
\newacronym{pdf}{PDF}{probability density function}
\newacronym{snr}{SNR}{signal-to-noise ratio}
\newacronym{bpsk}{BPSK}{binary phase shift keying}
\newacronym{psd}{PSD}{power spectral density}
\newacronym{mme}{MME}{maximum-minimum-eigenvalue}
\newacronym{idft}{IDFT}{inverse discrete Fourier transform}
\newacronym{awgn}{AWGN}{additive white Gaussian noise}

\usepackage{cleveref} 
\crefformat{equation}{(#1)} 
\Crefformat{equation}{(#1)} 
\let\ref\Cref    

\newcommand{\mean}[1]{\mathbb{E}[#1]}
\newcommand{\T}{^\text{T}} 
\renewcommand{\H}{^\text{H}} 
\newcommand{\liminfty}[1]{ \underset{#1\rightarrow \infty}{\operatorname{lim}} } 
\newcommand{\complexnum}{\ensuremath{\mathbb{C}}}  
\renewcommand{\vec}[1]{\ensuremath{\boldsymbol{\mathbf{\MakeLowercase{#1}}}}}
\newcommand{\mat}[1]{\ensuremath{\boldsymbol{\mathbf{\MakeUppercase{#1}}}}}
\newcommand{\hypzero}{\ensuremath{\mathcal{H}_0}} 
\newcommand{\hypone}{\ensuremath{\mathcal{H}_1}} 

\newcommand{\ie}[0]{i.\,e.}
\newcommand{\eg}[0]{e.\,g.}

\newcommand{\pmd}{P_\text{md}} 
\newcommand{\pfa}{P_\text{fa}} 
\newcommand{\hyptest}[0]{\overset{\scriptscriptstyle{\hypzero}}{\underset{\scriptscriptstyle{\hypone}}{\lessgtr}}} 
\newcommand{\threshold}{\gamma} 

\endlocaldefs

\begin{document}

\begin{frontmatter}

\begin{fmbox}
\dochead{Research}


\title{SNR-Walls in Eigenvalue-based Spectrum Sensing}


\author[
addressref={ti},
corref={ti},
email={bollig@ti.rwth-aachen.de}
]{\inits{AB}\fnm{Andreas} \snm{Bollig}}
\author[
addressref={ti},
corref={ti},
email={disch@ti.rwth-aachen.de}
]{\inits{CD}\fnm{Constantin} \snm{Disch}}
\author[
addressref={ti},
corref={ti},
email={arts@ti.rwth-aachen.de}
]{\inits{MA}\fnm{Martijn} \snm{Arts}}
\author[
addressref={ti},
corref={ti},
email={mathar@ti.rwth-aachen.de}
]{\inits{RM}\fnm{Rudolf} \snm{Mathar}}

\address[id=ti]{
	\orgname{Institute for Theoretical Information Technology, RWTH Aachen University},
	\street{Kopernikusstraße 16},
	\postcode{52074},
	\city{Aachen},
	\cny{Germany}
	}


\begin{artnotes}
\end{artnotes}

\end{fmbox}


\begin{abstractbox}

\begin{abstract} 
Various spectrum sensing approaches have been shown to suffer from a so-called SNR-wall, an SNR value below which a detector cannot perform robustly no matter how many observations are used.
Up to now, the eigenvalue-based \gls{mme} detector has been a notable exception.
For instance, the model uncertainty of imperfect knowledge of the receiver noise power, which is known to be responsible for the energy detector's fundamental limits, does not adversely affect the \gls{mme} detector's performance.
While \gls{awgn} is a standard assumption in wireless communications, it is not a reasonable one for the \gls{mme} detector.
In fact, in this work we prove that uncertainty in the amount of noise coloring does lead to an SNR-wall for the \gls{mme} detector.
We derive a lower bound on this SNR-wall and evaluate it for example scenarios.
The findings are supported by numerical simulations.	
	
%
\end{abstract}




\end{abstractbox}
%

\end{frontmatter}


\section{Introduction}
\label{sec:intro}
The recent years have seen an ever-growing demand for wireless spectrum not least due to the rise of the smartphone in consumer markets. However, as a result of the licensing policies of the preceding decades, most of the radio spectrum can only be used by fixed licensees, many of which only make use of their spectral bands at certain places or times.
To alleviate the need for spectrum and make better use of the given resources, \gls{osa} \cite{zhao_survey_2007} has emerged as a sub-field of cognitive radio.
In \gls{osa}, a spectral band can be used by unlicensed transceivers, so-called \glspl{su}, if they are certain that the licensee of the band, the so-called \gls{pu}, is not using it.

To make sure the occupancy status of a band is reliably detected, the \gls{su} \emph{senses} the spectrum before using it.
The goal of spectrum sensing is to decide between two hypotheses, the first of which states that the band of interest is free ($\hypzero$), such that the \gls{su} can make use of it. The second hypothesis states that the band is occupied ($\hypone$), in which case the \gls{su} should refrain from accessing the band.
The requirements spectrum sensing algorithms have to meet are quite demanding, \eg, the IEEE 802.22 standard for cognitive wireless regional area networks \cite{ieee_ieee_2011} states that an \gls{su} receiver should be able to reliably detect a \gls{pu} signal at a \gls{snr} of -22dB.
There are good reasons for these demanding requirements, like, \eg, the hidden terminal problem \cite[Ch. 14.3.3]{goldsmith_wireless_2005}.

A number of spectrum sensing algorithms have been proposed in the literature \cite{axell_spectrum_2012, yucek_survey_2009, zeng_review_2010}.
Under the ergodicity assumption, these algorithms are typically able to meet the above requirement, \ie, if enough samples are available, the \glspl{pdf} of the test statistics are well separable.
However, due to \emph{model uncertainties} caused by, \eg, colored or non-stationary background noise, non-ideal filters and imperfectly estimated parameters,  detection algorithms can exhibit so-called SNR-walls, \ie, SNR values, below which the detectors cannot robustly \cite{huber_robust_2011} decide between $\hypzero$ and $\hypone$.
The existence of SNR-walls has been established for the energy detector, the matched filter detector and the cyclostationarity detector \cite{sonnenschein_radiometric_1992, tandra_snr_2008, tandra_snr_2007}. 
A spectrum sensing algorithm that, to the best of our knowledge, has not been linked to the \gls{snr}-wall problem is the popular eigenvalue-based \gls{mme} detector.

The contributions of this paper are manifold.
We identify noise coloring as a model uncertainty adversely affecting the \gls{mme} detector.
Then we show that uncertainty in the amount of noise coloring leads to an \gls{snr}-wall for the \gls{mme} detector by using noise coloring to derive a lower bound on the \gls{snr}-wall of the \gls{mme} detector.
Finally, we support the analytical results with numerical simulations.

The rest of the paper is structured as follows.
In \Cref{sec:snrwall_signal_model}, we introduce our signal model alongside the test statistic of the \gls{mme} detector.
\Cref{snr_walls} formally defines the term \gls{snr}-wall, while \Cref{sec:noise-coloring} explains why colored noise is a reasonable assumption in wireless communication.
In \Cref{sec:snr-lower-bound} a lower bound on the \gls{snr}-wall for the \gls{mme} detector is derived.
Example scenarios with concrete values for the lower bound from \Cref{sec:snr-lower-bound} are given in \Cref{sec:examples_of_lower_bounds}.
A numerical evaluation of our results can be found in \Cref{sec:numerical_evaluation}, while \Cref{sec:snrwall-conclusion} concludes the work.

\section{Signal Model and MME Test Statistic}
\label{sec:snrwall_signal_model}
Consider the discrete time complex baseband signal $x(n)$ observed at a secondary system receiver, where $n$ denotes the discrete time index.
The task of a spectrum sensing algorithm is to decide between the following two hypotheses
\begin{align}
	\begin{split}
		\mathcal{H}_0 : x(n)&=\eta(n)\\
		\mathcal{H}_1 : x(n)&=s(n)+\eta(n),
	\end{split}
\end{align}
where $s(n)$ represents a \gls{pu} signal, while $\eta(n)$ stands for additive noise.
For the sake of simplicity, no channel fading effects are taken into account.

Throughout this work, the \gls{pu} is assumed to transmit a linearly modulated signal with symbol length $T_{\text{symbol}}$ exhibiting a rectangular pulse shape.
The signal is oversampled at the receiver with an integer oversampling rate given by $M=\frac{T_{\text{symbol}}}{T_{\text{sample}}}$, where $T_\text{sample}$ denotes the sampling period.
The decision whether the band under observation is free (\hypzero) or occupied (\hypone) is based on a block of $N$ samples. 
In the decision process, samples from $p$ different receivers are considered. 
The samples available at the fusion center at time instant $n$ are given by
\begin{equation}
	\label{sample_vector_composition}
	\begin{split}
		\vec x(n) = \left [x_1(n),x_1(n-1), \dotsc, x_1(n-Q),\right.\\
		\left. x_2(n), \dotsc, x_p(n), \dotsc, x_p(n-Q)\right]\T,
	\end{split}
\end{equation}
where the subscript indicates, which receiver the samples are from.
Each receiver contributes a consecutive set of $Q+1$ samples, where the quantity $Q + 1$ is the so called smoothing factor \cite{zeng_eigenvalue-based_2009}. 
The inclusion of samples from different receivers as well as samples from different points in time allows the \gls{mme} detector to exploit correlation from both domains in the detection process.
For simplicity, all receivers are assumed to experience the same \gls{snr}.
The vectors $\vec s(n)$ and $\vec \eta(n)$ are defined analogous to $\vec x(n)$, leading to the concise representation
\begin{equation}
	\vec x(n) = \vec s(n) + \vec \eta(n).	
\end{equation}

We consider a scenario with a single \gls{pu} transmitter and multiple \gls{su} receivers.
The receivers are assumed to be perfectly synchronized, \ie, $s_i(n)|_{i=1}^p = s(n)$.
Signal and noise are generated by mutually independent stationary random processes.
The \gls{pu} signal $s(n)$ is zero-mean, has variance $\sigma_s^2$ and its symbols are independent, \ie, $s(n)$ and $s(n + M)$ are independent and identically distributed (i.i.d.).
The receiver noise $\eta_i(n)$ is zero-mean and has variance $\sigma_\eta^2$ for all $i$.
The noise vector $\vec \eta(n)$ is distributed according to a circularly-symmetric complex Gaussian distribution, \ie, $\vec \eta(n) \sim \mathcal{CN}_{p(Q + 1)}(\vec 0, \mat R_{\vec \eta})$.

Considering the fact that both the signal and the noise are zero-mean, the statistical covariance matrices can be obtained as
\begin{align}\label{snr-wall-stat-cov-mats}
	\begin{split}
		\mat R_{\vec s} &= \mean{\vec{s}(n)\vec s(n)\H}\\
		\mat R_{\vec \eta} &= \mean{\vec \eta(n) \vec \eta(n)\H}\\
		\mat R_{\vec x} &= \mean{\vec x(n) \vec x(n)\H}\\
		&= \mat R_{\vec s} + \mat R_{\vec \eta},
	\end{split}
\end{align}
where $(\cdot)\H$ denotes the complex conjugate transpose.
We assume that the received signal $\vec x(n)$ is covariance ergodic \cite[pp. 531]{papoulis_probability_2002}, such that the sample covariance matrix
\begin{equation}\label{eq:sample_covariance_matrix}
	\hat{\mat R}_{\vec x}(N) = \frac 1{N-Q} \sum\limits_{n=Q}^{N-1}\vec x(n)\vec x(n)\H
\end{equation}
asymptotically converges to the statistical covariance matrix, \ie,
\begin{equation}
	\liminfty N\hat{\mat R}_{\vec x}(N) = \mat R_{\vec x}.
\end{equation}
The analytic derivations in this work target the well-known \gls{mme} detector \cite{zeng_eigenvalue-based_2009}. 
Its test statistic is composed of the eigenvalues of the received signal's sample covariance matrix. 
The test statistic and its accompanying decision rule are given by
\begin{equation}\label{hyp_test}
	\Gamma_\text{MME}(\vec x, N) = \frac{\lambda_\text{max}(\hat{\mat R}_{\vec x}(N))}{\lambda_\text{min}(\hat{\mat R}_{\vec x}(N))} \hyptest \threshold,
\end{equation}
where $\lambda_\text{max}(\cdot)$ and $\lambda_\text{min}(\cdot)$ denote the largest and smallest eigenvalue of a matrix respectively and $\threshold$ stands for the predefined decision threshold.
If $\Gamma_\text{MME}(\vec x, N) < \threshold$, we decide $\hypzero$, while for $\Gamma_\text{MME}(\vec x, N) \ge \threshold$ we decide $\hypone$.
Since both sample and statistical covariance matrices are positive-semidefinite and thus all of their eigenvalues are $\ge 0$, the test statistic $\Gamma_\text{MME}(\vec x, N)$ is always $\ge 1$.

\section{SNR Walls in Spectrum Sensing}
\label{snr_walls}
\Acrfull{awgn} is a standard assumption in wireless communications research and for many problems in the field, it is a reasonable one.
Indeed, a classical result from information theory states that additive Gaussian noise represents the worst case in point-to-point communication \cite{shomorony_worst-case_2013}, which makes it a fair choice for performance evaluation.
However, modeling the receiver noise as \gls{awgn} is only an approximation of reality and for eigenvalue-based spectrum sensing, where correlation in the received signal is the key to differentiability between the $\hypzero$ and the $\hypone$ case, it does not embody the worst case.
As will be shown in the subsequent sections, the assumption that the (Gaussian) noise samples are i.i.d. (and thus the noise is \emph{white}) is a crucial prerequisite for the \gls{mme} detector's optimal operation.
To take into consideration that different types of noise exist, we model $\eta(n)$ as having any distribution $W$ from a set of possible distributions $\mathcal W$, all of which have the variance $\sigma_\eta^2$.
The \gls{mme} detector is a general spectrum sensing algorithm in the sense that it is capable of detecting different kinds of signals. Thus, we do not assume a fixed signal type but instead only make the assumption that the \gls{pu} signal $s(n)$ has any distribution $S$ from the set of possible distributions $\mathcal S$, all of which have the variance $\sigma_s^2$.

Given the sets $\mathcal S$ and $\mathcal W$ with the variances $\sigma_s^2$ and $\sigma_\eta^2$ respectively, we can now define the \gls{snr} as
\begin{equation}
\text{SNR} = \frac{\sigma_s^2}{\sigma_\eta^2}.
\end{equation}
Further, we define the probability of false alarm and the probability of missed detection as
\begin{align}\begin{split}
\pfa(W, N) &= P(\Gamma_\text{MME}(\vec x, N) \ge \gamma\;|\;\hypzero, W),\\
\pmd(W, S, N) & = P(\Gamma_\text{MME}(\vec x, N) < \gamma\;|\;\hypone, W, S),
\end{split}\end{align}	respectively.
In conformity with the definition in \cite{tandra_snr_2008} (except that we do not consider a fading channel), we let a detector \emph{robustly} achieve a pair ($\pfa, \pmd$) consisting of a target false alarm probability $\pfa$ and a target missed detection probability $\pmd$ if it satisfies
\begin{align}\begin{split}
\underset{W \in \mathcal W}{\text{sup}} \pfa(W,N) &\le \pfa,\\
\underset{W \in \mathcal W, S \in \mathcal S}{\text{sup}} \pmd(W,S,N) &\le \pmd.
\end{split}\end{align}

The detector is called non-robust at a given SNR if at that SNR even with an arbitrarily high $N$, it cannot achieve any pair ($\pfa, \pmd$) on the support $\pfa \in [0,0.5], \pmd \in [0,0.5]$.
The SNR-wall is finally defined as 
\begin{align}\begin{split}
\text{SNR}_\text{wall} = \text{sup}\{\text{SNR}_t, \;\text{s.t. the detector is non-robust}\\\text{for all SNR} < \text{SNR}_t\}.
\end{split}\end{align}

For test statistics with symmetric \glspl{pdf}, a definition of non-robustness equivalent to the above is that a detector is non-robust iff the sets of means of the test statistic $\Gamma(\vec x, N)$ under the two hypotheses overlap \cite{tandra_snr_2008}.
However, the $\hypzero$ and $\hypone$ test statistic \glspl{pdf} of the \gls{mme} detector are not symmetric.
Thus, we use a third definition of non-robustness, which states that a detector is non-robust iff the sets of medians of the test statistic $\Gamma(\vec x, N)$ under the two hypotheses overlap.
This definition is equivalent to the first one, even for non-symmetric distributions. 
For symmetric distributions, it coincides with the second definition.
We make use of the third version of the definition of an \gls{snr}-wall when deriving a lower bound on the SNR-wall for the \gls{mme} detector, thus proving the existence of an SNR-wall for that detector, in \Cref{sec:snr-lower-bound}.

\section{Sources of noise and noise coloring}
\label{sec:noise-coloring}
As mentioned in \Cref{snr_walls}, \gls{awgn} can only be considered an approximation of the actual noise experienced at a radio receiver.
In this section we make the case for considering colored receiver noise, which is an assumption used in the subsequent sections to prove the existence of an SNR-wall for the \gls{mme} detector.

In \cite{zeng_eigenvalue-based_2009} it is assumed that the receiver noise before processing is white and that the only source of noise coloring is the use of a receive filter.
The receive filter is assumed to be known in advance and to be invertible, such that a pre-whitening filter can be used to re-whiten the received samples before including them in the computation of the test statistic.
On close inspection, it seems to be problematic to remove all coloring, since perfect filter design is hardly achievable, which makes employing an exact inverse of the receive filter seem impossible. 

Even if the coloring caused by the receiver architecture was perfectly reversible, the assumption that the external noise is white represents an oversimplification, which in the case of the \gls{mme} detector happens to be inappropriate, since  for this detector uncorrelated noise is a requirement for proper functioning. 
In reality, the external noise is a superposition of different kinds of noise from various sources and although the impact of external noise decreases for higher frequencies \cite{international_telecommunication_union_recommendation_2015}, at moderate frequencies such as the ones used for television broadcasting, external noise is present and should be considered. Note that the television bands are of utmost interest for spectrum sensing, \ie, many spectrum sensing algorithms originated in the context of the IEEE 802.22 standard \cite{ieee_ieee_2011}, which is concerned with communication in vacant television bands.

There are multiple different sources of realistic non-white external noise.
One example is galactic radiation noise.
The \gls{psd} of such noise is proportional to $\frac 1 {f^{2.7}}$ \cite{motchenbacher_low_1993}, where $f$ denotes the frequency, which makes it non-white.
Another example is man-made noise \cite{wagstaff_man-made_2003}.
The reason this kind of noise leads to correlation in the received samples under $\hypzero$ is twofold.
Firstly, it occurs in strong bursts that affect multiple receiver samples, which leads to time correlation. 
Secondly, when multiple receivers or multiple receive antennas are considered, the received noise is correlated in the case that multiple of them are in the range of the same man-made impulsive noise.
The origin of man-made noise lies in, \eg, unintended radiation from electrical machinery or power transmission lines. 
Furthermore, nearly every electronic device creates it and thus, impulsive man-made noise is an effect that needs to be taken into account.

Given the above reasoning, it can arguably be expected that even in the absence of a \gls{pu} signal in the observed band, a certain amount of correlation is present in the received samples.

\section{SNR-wall lower bound}
\label{sec:snr-lower-bound}
In this section, we derive a lower bound on the SNR-wall of the \gls{mme} detector, thus proving its existence.
More specifically, following the definition of the SNR-wall from \Cref{snr_walls}, we derive a lower bound on the SNR value below which the sets of medians of the test statistic $\Gamma(\vec x, N)$ under the two hypotheses ($\hypzero$ and $\hypone$) overlap even in the asymptotic case ($N \rightarrow \infty$).
To determine the location of the overlap we provide a lower bound on the test statistic under $\hypzero$ and an upper bound on the test statistic under $\hypone$.
The SNR below which the first of these two bounds has a higher value than the second one is the lower bound on the SNR-wall. 
Due to the fact that for $N \rightarrow \infty$, the \glspl{pdf} of the test statistic under the two hypotheses both become degenerate distributions, such that the \glspl{pdf} coincide with their medians, this is equivalent to the above definition of the SNR-wall. 

In our system model, the matrices $\mat R_{\vec s}$ and $\mat R_{\vec \eta}$ are no Toeplitz matrices.
The reason can be found in the composition of the vector $\vec x(n)$, \ie, the covariance matrices contain two types of correlation, time and receiver correlation (cf. \cref{sample_vector_composition}). 
Note, that this is the case despite the facts that firstly, due to our asymptotic approach, the matrices are statistical covariance matrices and secondly, the underlying noise processes are assumed to be stationary.
The correlation coefficients associated with the entry in the $i$-th row and $j$-th column of $\mat R_{\vec s}$ and $\mat R_{\vec \eta}$ are denoted by $\rho^s_{ij}$ and $\rho^\eta_{ij}$, respectively.
Bounds will be denoted by a bar above the respective symbol.

\subsection{Lower bound on the test statistic under $\hypzero$}

In the $\hypzero$ case, $\mat R_{\vec x} = \mat R_{\vec \eta}$, since no \gls{pu} signal exists.
For this case, we aim at finding a \emph{lower} bound $\bar \Gamma ^\text{asym, lo} _{\text{MME}, \hypzero}$ on the asymptotic test statistic, \ie,
\begin{align}
\frac{\bar \lambda _\text{max}^\text{lo}(\mat R_{\vec \eta})}{\bar \lambda _\text{min}^\text{up}(\mat R_{\vec \eta})} = \bar \Gamma ^\text{asym, lo} _{\text{MME}, \hypzero}
\le
\Gamma ^\text{asym} _{\text{MME}, \hypzero} = \frac{\lambda _\text{max}(\mat R_{\vec \eta})}{\lambda _\text{min}(\mat R_{\vec \eta})}.
\end{align}
To obtain the lower bound $\bar \Gamma ^\text{asym, lo} _{\text{MME}, \hypzero}$, we need to determine a lower bound on the largest eigenvalue ($\bar \lambda _\text{max}^\text{lo}(\mat R_{\vec \eta})$) and an upper bound on the smallest eigenvalue ($\bar \lambda _\text{min}^\text{up}(\mat R_{\vec \eta})$), \ie,
\begin{align}\begin{split}
\bar \lambda _\text{max}^\text{lo}(\mat R_{\vec \eta}) &\le \lambda _\text{max}(\mat R_{\vec \eta}),\\
\bar \lambda _\text{min}^\text{up}(\mat R_{\vec \eta}) &\ge \lambda _\text{min}(\mat R_{\vec \eta}).
\end{split}\end{align}
According to the Courant-Fischer theorem, the maximum and minimum eigenvalues $\lambda_\text{max}(\mat H)$ and $\lambda_\text{min}(\mat H)$ of a hermitian matrix $\mat H$ can be obtained by solving the following optimization problems \cite{gray_toeplitz_2006}
\begin{align}\label{courant_fischer_theorem}\begin{split}
\lambda_\text{max}(\mat H) &= \underset{\vec z : \vec z\H \vec z = 1}{\text{max}} \vec z\H \mat H \vec z,\\
\lambda_\text{min}(\mat H) &= \underset{\vec z : \vec z\H \vec z = 1}{\text{min}} \vec z\H \mat H \vec z.
\end{split}\end{align}
From \cref{courant_fischer_theorem} it directly follows, that
\begin{equation}
\lambda_\text{min}(\mat H) \le \vec z\H \mat H \vec z \le \lambda_\text{max}(\mat H),
\end{equation}
for an arbitrary normalized $\vec z$.
This means, that we can obtain the bounds  $\bar \lambda _\text{max}^\text{lo}(\mat R_{\vec \eta})$ and $\bar \lambda _\text{min}^\text{up}(\mat R_{\vec \eta})$ by simply evaluating $\bar \lambda _\text{max}^\text{lo}(\mat R_{\vec \eta}) = \vec z_1\H\mat R_{\vec \eta}\vec z_1$ and $\bar \lambda _\text{min}^\text{up}(\mat R_{\vec \eta}) = \vec z_2\H\mat R_{\vec \eta}\vec z_2$ using two arbitrary normalized vectors $\vec z_1$ and $\vec z_2$. 
However, an additional constraint we need to take care of when obtaining the bounds is that $\bar \lambda _\text{max}^\text{lo}(\mat R_{\vec \eta}) \ge \bar \lambda _\text{min}^\text{up}(\mat R_{\vec \eta})$ needs to hold.
In the following, we construct a set of vectors $\vec z_1$ and $\vec z_2$ that guarantees that this property is satisfied.

Given a specific covariance matrix $\mat R_{\vec \eta}$, the vectors are constructed as to extract its largest correlation coefficient $\rho^\eta_\text{max}$ with $|\rho^\eta_\text{max}| \ge |\rho^\eta_{ij}|\,\forall\,i,j$ with $i\ne j$.
For the below example, we assume that the largest correlation coefficient is located at the $k$-th column of the first row of the matrix.
Given the hermitian structure of covariance matrices and the assumed stationarity of the noise processes, this assumption can be made without loss of generality.
The considered covariance matrix has the following structure
\begin{equation}
\mat R_{\vec \eta} = \sigma_\eta^2
\begin{pmatrix}
1 			&\cdots 	& |\rho_{\text {max}}^{\eta}|e^{j\phi} 	& \cdots \\
\vdots 		&\ddots		& \cdots  					& \cdots \\
\ |\rho_{\text {max}}^{\eta}|e^{-j\phi}  & \cdots  & 1				& \cdots \\
\vdots 		&\vdots		& \vdots  					& \ddots \\  
\end{pmatrix}.
\end{equation}
The accompanying vectors $\vec z_1$ and $\vec z_2$ are given by
\begin{align}\begin{split}
\vec z_{1}&=\frac{1}{\sqrt{2}}[1,0,\dots,0,+e^{-j\phi},0,\dots,0]^{\text{T}},\\
\vec z_{2}&=\frac{1}{\sqrt{2}}[1,0,\dots,0,-e^{-j\phi},0,\dots,0]^{\text{T}},
\end{split}\end{align}
where  $\pm e^{-j\phi}$ is the $k$-th element of the respective vector, coinciding with the position of $|\rho_{\text {max}}^{\eta}|e^{\pm j\phi}$ in $\mat R_{\vec \eta}$.
We can now obtain
\begin{equation}\label{eigenval_bounds}\begin{split}
\bar \lambda _\text{max}^\text{lo}(\mat R_{\vec \eta}) = \vec z_1\H \mat R_{\vec \eta} \vec z_1 &= \sigma_\eta^2(1+ |\rho_{\text {max}}^{\eta}|),\\
\bar \lambda _\text{min}^\text{up}(\mat R_{\vec \eta}) = \vec z_2\H \mat R_{\vec \eta} \vec z_2 &= \sigma_\eta^2(1- |\rho_{\text {max}}^{\eta}|).
\end{split}\end{equation}
Note that the above argument can be made for an arbitrary position of $\rho_{\text {max}}^{\eta}$ by choosing $\vec z_1$ and $\vec z_2$ accordingly, leaving \cref{eigenval_bounds} unchanged.
Note also, that $\bar \lambda _\text{max}^\text{lo}(\mat R_{\vec \eta}) \ge \bar \lambda _\text{min}^\text{up}(\mat R_{\vec \eta})$ holds.
The lower bound on the test statistic under $\hypzero$ is now given by
\begin{equation}\label{lower_bound_h0}
\bar \Gamma ^\text{asym, lo} _{\text{MME}, \hypzero} = 
\frac{1+ |\rho_{\text {max}}^{\eta}|}{1- |\rho_{\text {max}}^{\eta}|},
\end{equation}
for $0 \le |\rho_{\text {max}}^{\eta}| < 1$.
For the case of complete noise correlation, \ie, $|\rho_{\text {max}}^{\eta}| = 1$ we get $0 \le\lambda _\text{min}(\mat R_{\vec \eta}) \le \bar \lambda _\text{min}^\text{up}(\mat R_{\vec \eta}) = 0$, where the first inequality is due to the general positive semidefiniteness of covariance matrices.
Since for $\lambda _\text{min}(\mat R_{\vec \eta}) = 0$ detection becomes impossible (cf. \cref{hyp_test}), we exclude the case of complete noise correlation.

\subsection{Upper bound on the test statistic under $\hypone$}
In the $\hypone$ case, $\mat R_{\vec x}= \mat R_{\vec s} + \mat R_{\vec \eta}$.
For this case, we aim at finding an \emph{upper} bound $\bar \Gamma ^\text{asym, up} _{\text{MME}, \hypone}$ on the asymptotic test statistic, \ie,
\begin{align}
\frac{\bar \lambda _\text{max}^\text{up}(\mat R_{\vec x})}{\bar \lambda _\text{min}^\text{lo}(\mat R_{\vec x})} = \bar \Gamma ^\text{asym, up} _{\text{MME}, \hypone}
\ge
\Gamma ^\text{asym} _{\text{MME}, \hypone} = \frac{\lambda _\text{max}(\mat R_{\vec x})}{\lambda _\text{min}(\mat R_{\vec x})}.
\end{align}
To obtain the upper bound $\bar \Gamma ^\text{asym, up} _{\text{MME}, \hypone}$, we need to determine an upper bound on the largest eigenvalue ($\bar \lambda _\text{max}^\text{up}(\mat R_{\vec x})$) and a lower bound on the smallest eigenvalue ($\bar \lambda _\text{min}^\text{lo}(\mat R_{\vec x})$), \ie,
\begin{align}\begin{split}
\bar \lambda _\text{max}^\text{up}(\mat R_{\vec x}) &\ge \lambda _\text{max}(\mat R_{\vec x}),\\
\bar \lambda _\text{min}^\text{lo}(\mat R_{\vec x}) &\le \lambda _\text{min}(\mat R_{\vec x}).
\end{split}\end{align}
Let $r_{ij}$ denote the entry of $\mat R_{\vec x}$ in the $i$-th row and $j$-th column.
Given our assumptions from \Cref{sec:snrwall_signal_model}, we get
\begin{align}\begin{split}
r_{ ii}&=\sigma_\eta^2+\sigma_s^2=(1+\operatorname{SNR})\sigma_\eta^2,\\
r_{ij}&=\sigma_\eta^2 \rho_{ij}^{\eta} +\sigma_s^2 \rho_{ij}^{s}=(\rho_{ij}^{\eta}+\operatorname{SNR} \rho_{ij}^{s})\sigma_\eta^2.
\end{split}
\end{align}
According to the Gershgorin circle theorem \cite{gerschgorin_uber_1931}, all eigenvalues $\lambda_k|_{k=1}^g$ of a matrix $\mat A \in \complexnum^{g\times g}$, with $g = p(Q+1)$, lie within the union of the circular disks
\begin{equation}
\label{gerschgorin_disks}
\{z \in \complexnum: |z-a_{ii}|\le R_i\},
\end{equation}
where
\begin{equation}
R_i=\sum_{\substack{j=1\\j\neq i}}^{g}{|a_{ij}|}.
\end{equation}
Since $\mat R_{\vec x}$ is hermitian, all of its eigenvalues are real and thus the disks from \cref{gerschgorin_disks} become intervals on the real axis.
The value of $r_{ii}$ is independent of $i$.
Thus, an upper bound on the maximum eigenvalue of $\mat R_{\vec x}$ can be obtained as
\begin{align}\label{h1_lambda_max_upper_bound}\begin{split}
\hspace{-.07cm}\bar \lambda _\text{max}^\text{up}(\mat R_{\vec x}) &= |r_{ii}| + \underset{i}{\text{max}}\; R_i\\
&=\sigma_\eta^2( \operatorname{SNR}+1+\underset{i}{\text{max}} \sum_{\substack{j=1\\j\neq i}}^{g}
|(\rho_{ij}^{\eta}+\operatorname{SNR} \rho_{ij}^{s})|)\\&\geq \lambda_{\text{max}}(\mat R_{\vec x}).
\end{split}\end{align}
In analogy to \cref{h1_lambda_max_upper_bound}, a lower bound on the minimum eigenvalue of $\mat R_{\vec x}$ can be obtained as
\begin{align}\label{h1_lambda_min_lower_bound}\begin{split}
\hspace{-.07cm}\bar \lambda _\text{min}^\text{lo}(\mat R_{\vec x}) &= |r_{ii}| - \underset{i}{\text{max}}\; R_i\\
&=\sigma_\eta^2( \operatorname{SNR}+1-\underset{i}{\text{max}} \sum_{\substack{j=1\\j\neq i}}^{g}
|(\rho_{ij}^{\eta}+\operatorname{SNR} \rho_{ij}^{s})|)\\&\le \lambda_{\text{min}}(\mat R_{\vec x}).
\end{split}\end{align}
However, for \cref{h1_lambda_min_lower_bound} to be a valid bound in terms of the test statistic, we need to introduce an extra constraint.
We need to make sure that $\bar \lambda _\text{min}^\text{lo}(\mat R_{\vec x}) > 0$, \ie, $|r_{ii}|-\underset{i}{\text{max}}~R_i>0$, which leads to the constraint
\begin{align}
1+\operatorname{SNR}>\underset{i}{\text{max}}\sum_{\substack{j=1\\j\neq i}}^{g} |(\rho_{ij}^{\eta}+\operatorname{SNR} \rho_{ij}^{s})|. \label{eq:conts_min_lambd2}
\end{align}
Combining \cref{h1_lambda_max_upper_bound} and \cref{h1_lambda_min_lower_bound} finally provides the upper bound on the asymptotic test statistic under $\hypone$ given by
\begin{align}
\bar{\Gamma}_{\text{MME},\mathcal{H}_1}^{\text {asym, up}}=\frac{\operatorname{SNR}+1+\underset{i}{\text{max}} \sum\limits_{\substack{j=1\\j\neq i}}^{g}
	|(\rho_{ij}^{\eta}+\operatorname{SNR} \rho_{ij}^{s})|}{\operatorname{SNR}+1-\underset{i}{\text{max}} \sum\limits_{\substack{j=1\\j\neq i}}^{g}|(\rho_{ij}^{\eta}+\operatorname{SNR} \rho_{ij}^{s})|}.
\label{eq:bound_gamma_asym_H1}
\end{align}

\subsection{Lower bound on the SNR-wall}

Combining \cref{lower_bound_h0} and \cref{eq:bound_gamma_asym_H1}, we can say that the \gls{mme} detector is non-robust under the condition given by \cref{eq:conts_min_lambd2} and given that $\rho^\eta_\text{max} <1$ when $\bar{\Gamma}_{\text{MME},\mathcal{H}_1}^{\text {asym, up}} \le \bar \Gamma ^\text{asym, lo} _{\text{MME}, \hypzero}$, \ie,
\begin{align}
\hspace{-.05cm}\frac{\operatorname{SNR}+1+\underset{i}{\text{max}} \sum\limits_{\substack{j=1\\j\neq i}}^{g}|(\rho_{ij}^{\eta}+\operatorname{SNR} \rho_{ij}^{s})|}{\operatorname{SNR}+1-\underset{i}{\text{max}} \sum\limits_{\substack{j=1\\j\neq i}}^{g}|(\rho_{ij}^{\eta}+\operatorname{\operatorname{SNR}} \rho_{ij}^{s})|}
\le \frac{1+|\varrho_{\text {max}}^{\eta}|}{1-|\varrho_{\text {max}}^{\eta}|}.
\label{eq:SNR_wall_bound}
\end{align}
Note, that the correlation coefficients in the $\hypzero$ case are now denoted by $\varrho$ instead of $\rho$ to facilitate the distinction between the noise correlation in the $\hypzero$ and the $\hypone$ case.
For the interpretation of \cref{eq:SNR_wall_bound} it is important to note, that $\rho_{ij}^s \ge 0 \;\forall\; i,j$, \ie, the signal correlation coefficients never become negative. 
This follows from \cref{sample_vector_composition}, \cref{snr-wall-stat-cov-mats} and the assumption that consecutive symbols are independent.

\section{Examples of the lower bound on the SNR-wall}
\label{sec:examples_of_lower_bounds}
Inequality \cref{eq:SNR_wall_bound} is quite involved and does not allow for easy interpretation.
Since our goal is to prove the existence of an \gls{snr}-wall, we continue our investigation with examples that simplify \cref{eq:SNR_wall_bound}, facilitating interpretation.
We consider the case, where under $\hypone$ no noise correlation exists, \ie, $\rho_{ij}^\eta = 0\;\forall\;i,j$ except $i=j$, while under $\hypzero$, the noise is correlated, \ie, $\exists\; (i,j)$ with $i \ne j$ for which $\varrho_{ij}^\eta \ne 0$.
This case occurs when the sources of noise coloring in the vicinity of the sensor are only present at certain times or if the sensor is used at different locations.
Considering uncorrelated noise under $\hypone$, \cref{eq:conts_min_lambd2} can be simplified as follows
\begin{align}\begin{split}
&1+\operatorname{SNR}>\underset{i}{\text{max}}\sum_{\substack{j=1\\j\neq i}}^{g} |(\rho_{ij}^{\eta}+\operatorname{SNR} \rho_{ij}^{s})|\\
\Leftrightarrow &1+\operatorname{SNR}>\operatorname{SNR}\cdot \underset{i}{\text{max}}\sum_{\substack{j=1\\j\neq i}}^{g} |\rho_{ij}^{s}|\\
\Leftrightarrow &\operatorname{SNR}<\frac{1}{\kappa_{\text {max}}-1},\label{eq:conts_min_lambd2_simp}
\end{split}\end{align}
where
\begin{equation}
\kappa_{\text {max}}=\underset{i}{\text{max}}\sum_{\substack{j=1\\j\neq i}}^{g}|\rho_{ij}^{s}|.
\end{equation}
This means, that the higher the correlation in the signal samples, the lower the \gls{snr} for which our lower bound is defined.
For $\kappa_{\text {max}} < 1$, the condition is never satisfied.
In this case, we have to fall back to zero as a lower bound for $\lambda_\text{min}$, which is guaranteed by the properties of covariance matrices.
This however leads to the test statistic under $\hypzero$ taking the value infinity, which again rules out the possibility of giving a lower bound for an \gls{snr}-wall.
For a more concise notation, let 
\begin{equation}
\alpha_\text{max} = \bar \Gamma ^\text{asym, lo} _{\text{MME}, \hypzero} = \frac{1+|\varrho_{\text {max}}^{\eta}|}{1-|\varrho_{\text {max}}^{\eta}|}.
\end{equation}
By assuming a minimal amount of noise coloring under $\hypzero$ and excluding the case of complete noise correlation, we restricted the support of the largest noise correlation coefficient, such that $\varrho_{\text {max}} \in (0, 1)$.
As a consequence, it holds that $\alpha_\text{max} > 1$.
Using the definitions of $\kappa_{\text {max}}$ and $\alpha_\text{max}$, as well as the assumption that under $\hypone$ the noise is uncorrelated, \cref{eq:SNR_wall_bound} becomes
\begin{align}
\frac{\operatorname{SNR}+1+\kappa_{\text {max}}\operatorname{SNR}}{\operatorname{SNR}+1-\kappa_{\text {max}}\operatorname{SNR}}\le \alpha_{\text {max}},
\end{align} or equivalently
\begin{align}
\operatorname{SNR} \le \frac{\alpha_{\text {max}}-1}{1+\kappa_{\text {max}}+\alpha_{\text {max}}(\kappa_{\text {max}}-1)}.
\label{eq:SNR_wall_bound_simp}
\end{align}

In order to obtain concrete numbers for the bound, we will look at more specific examples in the following.

\subsection{Receiver correlation ($Q=0$, $p \ge 2$)}
\label{subsec:rec_corr}
In this example we consider a $p$-receiver setup with perfect signal correlation, \ie, $\rho_{ij}^s = 1\;\forall\;i, j$.
The maximum signal correlation in this case is $\kappa_{\text{max}} = p - 1$ and thus the condition \cref{eq:conts_min_lambd2_simp} becomes $\text{SNR} < \frac 1 {p-2}$.
If the condition is satisfied, we can say that the \gls{mme} detector becomes non-robust for
\begin{equation} \operatorname{SNR} \le \frac{\alpha_{\text {max}}-1}{p+\alpha_{\text {max}}(p-2)}.
\label{eq:snr_wall_rec_cor}
\end{equation}
For $p=2$ and a maximum noise correlation of $\varrho_\text{max}^\eta = 0.05$, which we consider to be moderate noise coloring, we arrive at a lower bound of $\text{SNR} = 0.052632 = -12.788\text{dB}$, which is considerably far away from $-22\text{dB}$ (cf. \Cref{sec:intro}).
This example is simple enough for us to obtain the actual statistical covariance matrices for an evaluation of the bound's tightness.
They are given by
\begin{align}\begin{split}
\mat R_{\vec x}^{\mathcal{H}_0}&=\mat R_{\vec \eta}^{\mathcal{H}_0}=\sigma_\eta^2
\begin{pmatrix}
1 			&0.05\\ 	
0.05 		&1		
\end{pmatrix}, \\
\mat R_{\vec x}^{\mathcal{H}_1}&=
\begin{pmatrix}
\sigma_\eta^2+\sigma_s^2 			&\sigma_s^2\\ 	
\sigma_s^2 						&\sigma_\eta^2+\sigma_s^2		
\end{pmatrix} \\  
&=\sigma_\eta^2
\begin{pmatrix}
1+\operatorname{SNR} 			&\operatorname{SNR}\\ 	
\operatorname{SNR} 				&1+\operatorname{SNR}		
\end{pmatrix}.
\end{split}\end{align}
With the above covariance matrices, the asymptotic test statistics evaluate to $\Gamma_{\mathcal{H}_0}^{\text{asym}}=1.10503$ and $\Gamma_{\mathcal{H}_1}^{\text {asym}}=1+2\operatorname{SNR}$.
This means, that for an \gls{snr} below $0.052632 = -12.788\text{dB}$, $\Gamma_{\mathcal{H}_1}^{\text{asym}} < \Gamma_{\mathcal{H}_0}^{\text{asym}}$, such that the detector becomes non-robust, \ie, in this special case the bound is tight.

\subsection{Time correlation ($p=1$, $Q\ge 1$)}
\label{subsec:time_corr}
To complement the receiver correlation example, we next investigate a case with only one receiver that examines the correlation of the received samples over time.
In order to again obtain a bound via \cref{eq:SNR_wall_bound_simp}, the value of $\kappa_{\text {max}}$ needs to be determined.
Given an oversampling factor $M$ and the independence of consecutive symbols, the autocorrelation function of the \gls{pu} signal can be obtained as
\begin{equation}
\mean{s^*(n)s(n\pm k)}=\begin{cases} \sigma_s^2(1-\frac{k}{M}) &\mbox{if } |k|<M \\ 0 & \mbox{else.}\end{cases}
\end{equation}
In our case, the row with the maximum off-diagonal correlation sum in $\mat R_{\vec s}$ is the \mbox{$\left\lfloor \frac{g + 1} 2 \right\rfloor$-th} one, where $g = Q + 1$ is the number of rows of $\mat R_{\vec s}$ and $\lfloor \cdot\rfloor$ denotes the floor operation.
This is illustrated in the following example for $Q=3$ (assuming $M \ge 3$)
\begin{align}
\mat R_{\vec s}=
\sigma_s^2
\begin{pmatrix}
1			&1-\frac{1}{M}	&1-\frac{2}{M}	&1-\frac{3}{M}\\ 	
1-\frac{1}{M}	&1			&1-\frac{1}{M}	&1-\frac{2}{M}\\
1-\frac{2}{M}	&1-\frac{1}{M}	&1			&1-\frac{1}{M}\\
1-\frac{3}{M}	&1-\frac{2}{M}	&1-\frac{1}{M}	&1	
\end{pmatrix}.
\end{align}
The further away from the diagonal, the smaller the value.
Thus, the middle row has the highest sum.

In order to obtain $\kappa_{\text {max}}$, three cases have to be distinguished.
For $\left\lceil\frac Q 2\right\rceil < M$ and Q even, we get 
\begin{align}\begin{split}
\kappa_{\text {max}} &=2\sum_{j=1}^{\frac{Q}{2}}{\left(1-\frac j M\right)}\\
&=Q-\frac{Q^2 + 2Q}{4M},
\end{split}\end{align}
for $\left\lceil\frac Q 2\right\rceil < M$ and Q odd, we get
\begin{align}\begin{split}
\kappa_{\text {max}}&=2 \sum_{j=1}^{\frac{Q-1}{2}}{\left(1-\frac j M\right)}+\left(1-\frac{Q+1}{2M}\right)\\
&=Q-\frac{(Q+1)^2}{4M},
\label{eq:kappa_max_time_Q_odd}
\end{split}\end{align}
and for $\left\lceil\frac Q 2\right\rceil \ge M$, we get
\begin{equation}
\kappa_{\text {max}}=2 \sum_{j=1}^{M-1}{\left(1-\frac j M\right)}=M-1,
\end{equation}
where $\lceil \cdot \rceil$ denotes the ceiling operation.
In this example we model the noise as a stationary auto-regressive process of order one (AR(1)).
It is supposed to mimic white external noise that has undergone filtering by a low-pass receive filter.
The noise process is given by
\begin{equation}
\eta(n) = 0.1 \eta(n - 1) + \epsilon(n),
\end{equation}
where $\epsilon(n)$ denotes an i.i.d. complex Gaussian random process with mean zero and variance $0.99$.
It is independent of $\eta(n - 1)$, and has independent real and imaginary parts and $\mean {|\eta [n]| ^2}=1$.
\Cref{fig:psd_ar_1} shows the noise process's \gls{psd} to illustrate its characteristics.
We consider the case of $Q=3$ and $M=4$, where the choice of $Q$ leads to the noise covariance matrix
\begin{equation}
\mat R_{\vec\eta}= \begin{pmatrix}
1			&0.1			&0.1^2			&0.1^3\\ 	
0.1		&1				&0.1			&0.1^2\\
0.1^2		&0.1			&1				&0.1\\
0.1^3		&0.1^2			&0.1			&1		
\end{pmatrix}
\end{equation}
and $\kappa_\text{max} = 2$ (cf. \cref{eq:kappa_max_time_Q_odd}).
Using \cref{eq:SNR_wall_bound_simp}, we again get a lower bound for the \gls{snr}-wall of $-12.788$dB.

\subsection{Time and receiver correlation ($p>1, Q>0$)}
As a final example, we consider the case where both, time and receiver correlation, are exploited.
Given our model assumption that the signal strength is equal for all receivers, we can combine the $\kappa_\text{max}$ terms derived in the preceding subsections to obtain
\begin{equation}
\kappa_{\text {max}}=p-1+p\cdot\kappa_{\text {max,time}}
\label{eq:kappa_max_time__rec_cor},
\end{equation}
where $\kappa_{\text {max,time}}$ denotes the $\kappa_{\text {max}}$ term for time correlation.

\section{Numerical Evaluation}
\label{sec:numerical_evaluation}
In this section we provide numerical results corresponding to the examples given in \Cref{subsec:rec_corr} and \Cref{subsec:time_corr}.
The parameters used in the simulations can be found in \Cref{tab:colnoise_sim_par}.

For the $\hypone$ case we generate white Gaussian noise and an oversampled \gls{bpsk} signal with a rectangular pulse shape and symbols that are independent of each other.
For the $\hypzero$ case we generate colored noise. 
The different noise types used for $\hypzero$ and $\hypone$ represent the model uncertainty that has to be taken into account when designing spectrum sensing algorithms.
When generating colored noise, our aim is to create a stationary, sampled Gaussian process with a predefined covariance matrix.

\subsection{Receiver correlation}

In the receiver correlation case, we use a matrix multiplication approach for generating colored noise.
We start by generating the $p$-dimensional white Gaussian noise sample vector $\vec w(n) \sim \mathcal N(\vec 0, \mat R_{\vec w})$ for all time instances $n$, where $\mat R_{\vec w} = \mat I_{p \times p}$, while $p$ is the number of receivers.
That is, the vector $\vec w(n)$  is generated according to a $p$-dimensional zero-mean Gaussian distribution with the identity matrix as its covariance matrix.
The colored noise vector $\vec \eta(n)$, which is experienced at the receiver in the $\hypzero$ case, is subsequently obtained as $\vec \eta(n) = \mat A \vec w(n)$, where the matrix $\mat A \in \complexnum^{p\times p}$ needs to be chosen such that the covariance matrix of $\vec \eta(n)$ equals the predefined $\mat R_{\vec \eta}$.
This approach leads to the desired result due to the fact that linear combinations of Gaussian random variables are again distributed according to a Gaussian distribution.
It is well known that the covariance matrix of $\mat A \vec w(n)$ is given by $\mat A \H \mat R_{\vec w} \mat A$.
Thus, the matrix $\mat A$ can be easily obtained by computing the Cholesky decomposition $\mat R_{\vec \eta} = \mat A\H\mat A$.

\Cref{fig:pdf_plot_coloured_noise_rec_cor} shows the histograms for the \gls{mme} test statistic $\Gamma_\text{MME}$ in the $\hypzero$ case (black) and the $\hypone$ case (colored).
For the $\hypone$ case, the test statistic histograms for different \glspl{snr} are shown.
It can be observed that the estimation variance of the test statistic decreases with an increasing number of samples $N$, which is to be expected since asymptotically the sample covariance matrix converges to the statistical covariance matrix.
What we can also see is that the mean of the estimated test statistic changes for an increasing $N$.
This is a testimony to the biasedness of the estimator \cref{hyp_test}.
Recall, that the lower bound on the \gls{snr}-wall for this scenario has been derived to be $-12.788$dB.
The simulation results confirm this bound.
Indeed, when the \gls{snr} drops below the derived bound, the medians of the test statistics under $\hypone$ are below the median of the test statistic under $\hypzero$ for all inspected $N$, making the detector non-robust by definition.

\subsection{Time correlation}

Since in the previous example, the noise correlation only exists between the noise processes of different receivers but not between noise samples taken at a single receiver at different times, the matrix $\mat A$, which is used to color the noise, has a small dimension.
This makes the matrix multiplication approach feasible for the receiver correlation example.
To use the same method in the time correlation example, a coloring matrix of size $N \times N$ would be necessary, where in our simulations, $N$ takes on values of up to $10^6$.
This renders the matrix multiplication approach infeasible for the current example.
Thus, a different approach for generating colored noise has to be taken.
First, we generate an autocorrelation with a real-valued \gls{psd} from an autoregressive model of order one.
We then generate an $N$-dimensional vector distributed according to a zero-mean, unit-variance, complex, white Gaussian distribution, which serves as a frequency-domain noise basis.
Its \gls{psd} is subsequently scaled by the \gls{psd} of the autocorrelation, after which it is transformed to the time-domain via the \gls{idft} and scaled to variance $\sigma_\eta^2$.
Here again, we use the fact, that a linear combination of Gaussian random variables is also distributed according to a Gaussian distribution.

Recall, that the lower bound on the \gls{snr}-wall for this scenario has been derived to be $-12.788$dB.
According to the numerical results, in this scenario the \gls{mme} detector exhibits an \gls{snr}-wall between $-8$dB and $-9$dB.
While the lower bound cannot be called very tight for this example, it nevertheless proves the existence of an \gls{snr}-wall, which is guaranteed to be much higher than the desired $-22$dB.

\section{Conclusion}
\label{sec:snrwall-conclusion}
In this work, we have proven the existence of an \gls{snr}-wall for the eigenvalue-based \gls{mme} spectrum sensing algorithm.
The \gls{snr}-wall is caused by uncertainty about the coloring of the receiver noise.
A lower bound on the \gls{snr}-wall is derived and is complemented by time and receiver correlation examples.
For the examples, we give concrete \gls{snr} values, below which the \gls{mme} detector is non-robust.
Finally, numerical results supporting the analytical results are provided.

One possible direction for future work is the derivation of tighter bounds on the \gls{snr}-wall.
Also, it would be of great value if the results could be generalized to arbitrary eigenvalue-based spectrum sensing algorithms.


\begin{backmatter}

%
%
%
\section*{Author information}
A. Bollig, C. Disch, M. Arts and R. Mathar are with the Institute for Theoretical Information Technology,
RWTH Aachen University, D-52074 Aachen, Germany. E-mail: \{bollig, disch, arts, mathar\}@ti.rwth-aachen.de.

\section*{Funding}
This work was partly supported by the Deutsche Forschungsgemeinschaft (DFG) project CoCoSa (grant MA 1184/26-1).  

\section*{Competing interests}
The authors declare that they have no competing interests.


\bibliographystyle{bmc-mathphys} 
\bibliography{snr_wall_paper}      


\begin{thebibliography}{18}
\ifx \bisbn   \undefined \def \bisbn  #1{ISBN #1}\fi
\ifx \binits  \undefined \def \binits#1{#1}\fi
\ifx \bauthor  \undefined \def \bauthor#1{#1}\fi
\ifx \batitle  \undefined \def \batitle#1{#1}\fi
\ifx \bjtitle  \undefined \def \bjtitle#1{#1}\fi
\ifx \bvolume  \undefined \def \bvolume#1{\textbf{#1}}\fi
\ifx \byear  \undefined \def \byear#1{#1}\fi
\ifx \bissue  \undefined \def \bissue#1{#1}\fi
\ifx \bfpage  \undefined \def \bfpage#1{#1}\fi
\ifx \blpage  \undefined \def \blpage #1{#1}\fi
\ifx \burl  \undefined \def \burl#1{\textsf{#1}}\fi
\ifx \doiurl  \undefined \def \doiurl#1{\textsf{#1}}\fi
\ifx \betal  \undefined \def \betal{\textit{et al.}}\fi
\ifx \binstitute  \undefined \def \binstitute#1{#1}\fi
\ifx \binstitutionaled  \undefined \def \binstitutionaled#1{#1}\fi
\ifx \bctitle  \undefined \def \bctitle#1{#1}\fi
\ifx \beditor  \undefined \def \beditor#1{#1}\fi
\ifx \bpublisher  \undefined \def \bpublisher#1{#1}\fi
\ifx \bbtitle  \undefined \def \bbtitle#1{#1}\fi
\ifx \bedition  \undefined \def \bedition#1{#1}\fi
\ifx \bseriesno  \undefined \def \bseriesno#1{#1}\fi
\ifx \blocation  \undefined \def \blocation#1{#1}\fi
\ifx \bsertitle  \undefined \def \bsertitle#1{#1}\fi
\ifx \bsnm \undefined \def \bsnm#1{#1}\fi
\ifx \bsuffix \undefined \def \bsuffix#1{#1}\fi
\ifx \bparticle \undefined \def \bparticle#1{#1}\fi
\ifx \barticle \undefined \def \barticle#1{#1}\fi
\ifx \bconfdate \undefined \def \bconfdate #1{#1}\fi
\ifx \botherref \undefined \def \botherref #1{#1}\fi
\ifx \url \undefined \def \url#1{\textsf{#1}}\fi
\ifx \bchapter \undefined \def \bchapter#1{#1}\fi
\ifx \bbook \undefined \def \bbook#1{#1}\fi
\ifx \bcomment \undefined \def \bcomment#1{#1}\fi
\ifx \oauthor \undefined \def \oauthor#1{#1}\fi
\ifx \citeauthoryear \undefined \def \citeauthoryear#1{#1}\fi
\ifx \endbibitem  \undefined \def \endbibitem {}\fi
\ifx \bconflocation  \undefined \def \bconflocation#1{#1}\fi
\ifx \arxivurl  \undefined \def \arxivurl#1{\textsf{#1}}\fi
\csname PreBibitemsHook\endcsname

\bibitem{zhao_survey_2007}
\begin{barticle}
\bauthor{\bsnm{Zhao}, \binits{Q.}},
\bauthor{\bsnm{Sadler}, \binits{B.M.}}:
\batitle{A {Survey} of {Dynamic} {Spectrum} {Access}}.
\bjtitle{IEEE Signal Processing Magazine}
\bvolume{24}(\bissue{3}),
\bfpage{79}--\blpage{89}
(\byear{2007})
\end{barticle}
\endbibitem

\bibitem{ieee_ieee_2011}
\begin{botherref}
\oauthor{\bsnm{IEEE}}:
{IEEE} {Standard} for {Information} {Technology}–{Telecommunications} and
  information exchange between systems; {Wireless} {Regional} {Area} {Networks}
  ({WRAN})–{Specific} requirements {Part} 22: {Cognitive} {Wireless} {RAN}
  {Medium} {Access} {Control} ({MAC}) and {Physical} {Layer} ({PHY})
  {Specifications}: {Policies} and {Procedures} for {Operation} in the {TV}
  {Bands}.
IEEE Std 802.22-2011
(2011)
\end{botherref}
\endbibitem

\bibitem{goldsmith_wireless_2005}
\begin{bbook}
\bauthor{\bsnm{Goldsmith}, \binits{A.}}:
\bbtitle{Wireless {Communications}}.
\bpublisher{Cambridge University Press},
\blocation{Cambridge}
(\byear{2005})
\end{bbook}
\endbibitem

\bibitem{axell_spectrum_2012}
\begin{barticle}
\bauthor{\bsnm{Axell}, \binits{E.}},
\bauthor{\bsnm{Leus}, \binits{G.}},
\bauthor{\bsnm{Larsson}, \binits{E.G.}},
\bauthor{\bsnm{Poor}, \binits{H.V.}}:
\batitle{Spectrum {Sensing} for {Cognitive} {Radio} : {State}-of-the-{Art} and
  {Recent} {Advances}}.
\bjtitle{IEEE Signal Processing Magazine}
\bvolume{29}(\bissue{3}),
\bfpage{101}--\blpage{116}
(\byear{2012})
\end{barticle}
\endbibitem

\bibitem{yucek_survey_2009}
\begin{barticle}
\bauthor{\bsnm{Yücek}, \binits{T.}},
\bauthor{\bsnm{Arslan}, \binits{H.}}:
\batitle{A {Survey} of {Spectrum} {Sensing} {Algorithms} for {Cognitive}
  {Radio} {Applications}}.
\bjtitle{IEEE Communications Surveys Tutorials}
\bvolume{11}(\bissue{1}),
\bfpage{116}--\blpage{130}
(\byear{2009})
\end{barticle}
\endbibitem

\bibitem{zeng_review_2010}
\begin{botherref}
\oauthor{\bsnm{Zeng}, \binits{Y.}},
\oauthor{\bsnm{Liang}, \binits{Y.-C.}},
\oauthor{\bsnm{Hoang}, \binits{A.T.}},
\oauthor{\bsnm{Zhang}, \binits{R.}}:
A {Review} on {Spectrum} {Sensing} for {Cognitive} {Radio}: {Challenges} and
  {Solutions}.
EURASIP Journal on Advances in Signal Processing
\textbf{2010}
(2010)
\end{botherref}
\endbibitem

\bibitem{huber_robust_2011}
\begin{bbook}
\bauthor{\bsnm{Huber}, \binits{P.J.}}:
\bbtitle{Robust Statistics}.
\bpublisher{Springer},
\blocation{Berlin}
(\byear{2011})
\end{bbook}
\endbibitem

\bibitem{sonnenschein_radiometric_1992}
\begin{barticle}
\bauthor{\bsnm{Sonnenschein}, \binits{A.}},
\bauthor{\bsnm{Fishman}, \binits{P.M.}}:
\batitle{Radiometric detection of spread-spectrum signals in noise of uncertain
  power}.
\bjtitle{IEEE Transactions on Aerospace and Electronic Systems}
\bvolume{28}(\bissue{3}),
\bfpage{654}--\blpage{660}
(\byear{1992})
\end{barticle}
\endbibitem

\bibitem{tandra_snr_2008}
\begin{barticle}
\bauthor{\bsnm{Tandra}, \binits{R.}},
\bauthor{\bsnm{Sahai}, \binits{A.}}:
\batitle{{SNR} {Walls} for {Signal} {Detection}}.
\bjtitle{IEEE Journal of Selected Topics in Signal Processing}
\bvolume{2}(\bissue{1}),
\bfpage{4}--\blpage{17}
(\byear{2008})
\end{barticle}
\endbibitem

\bibitem{tandra_snr_2007}
\begin{bchapter}
\bauthor{\bsnm{Tandra}, \binits{R.}},
\bauthor{\bsnm{Sahai}, \binits{A.}}:
\bctitle{{SNR} {Walls} for {Feature} {Detectors}}.
In: \bbtitle{{IEEE} {International} {Symposium} on {New} {Frontiers} in
  {Dynamic} {Spectrum} {Access} {Networks} ({DySPAN})},
pp. \bfpage{559}--\blpage{570}
(\byear{2007})
\end{bchapter}
\endbibitem

\bibitem{zeng_eigenvalue-based_2009}
\begin{barticle}
\bauthor{\bsnm{Zeng}, \binits{Y.}},
\bauthor{\bsnm{Liang}, \binits{Y.-C.}}:
\batitle{Eigenvalue-based spectrum sensing algorithms for cognitive radio}.
\bjtitle{IEEE Transactions on Communications}
\bvolume{57}(\bissue{6}),
\bfpage{1784}--\blpage{1793}
(\byear{2009})
\end{barticle}
\endbibitem

\bibitem{papoulis_probability_2002}
\begin{bbook}
\bauthor{\bsnm{Papoulis}, \binits{A.}},
\bauthor{\bsnm{Pillai}, \binits{S.U.}}:
\bbtitle{Probability, Random Variables, and Stochastic Processes},
\bedition{4}th edn.
\bpublisher{McGraw-Hill},
\blocation{New York City}
(\byear{2002})
\end{bbook}
\endbibitem

\bibitem{shomorony_worst-case_2013}
\begin{barticle}
\bauthor{\bsnm{Shomorony}, \binits{I.}},
\bauthor{\bsnm{Avestimehr}, \binits{A.S.}}:
\batitle{Worst-{Case} {Additive} {Noise} in {Wireless} {Networks}}.
\bjtitle{IEEE Transactions on Information Theory}
\bvolume{59}(\bissue{6}),
\bfpage{3833}--\blpage{3847}
(\byear{2013})
\end{barticle}
\endbibitem

\bibitem{international_telecommunication_union_recommendation_2015}
\begin{botherref}
\oauthor{\bsnm{{International Telecommunication Union}}}:
Recommendation {P}.372-12 : {Radio} noise.
Technical report
(July 2015).
\url{https://www.itu.int/rec/R-REC-P.372-12-201507-I/en}
Accessed 2015-11-23
\end{botherref}
\endbibitem

\bibitem{motchenbacher_low_1993}
\begin{bbook}
\bauthor{\bsnm{Motchenbacher}, \binits{C.D.}},
\bauthor{\bsnm{Connelly}, \binits{J.A.}}:
\bbtitle{Low Noise Electronic System Design}.
\bpublisher{Wiley},
\blocation{Hoboken}
(\byear{1993})
\end{bbook}
\endbibitem

\bibitem{wagstaff_man-made_2003}
\begin{botherref}
\oauthor{\bsnm{Wagstaff}, \binits{A.}},
\oauthor{\bsnm{Merricks}, \binits{N.}}:
Man-{Made} {Noise} {Measurement} {Programme} ({AY}4119) - {Final} {Report}.
Technical Report Issue 2,
Mass Consultants Limited
(September 2003)
\end{botherref}
\endbibitem

\bibitem{gray_toeplitz_2006}
\begin{bbook}
\bauthor{\bsnm{Gray}, \binits{R.M.}}:
\bbtitle{Toeplitz and Circulant Matrices: {A} Review}.
\bpublisher{Now Publishers},
\blocation{Boston, Delft}
(\byear{2006})
\end{bbook}
\endbibitem

\bibitem{gerschgorin_uber_1931}
\begin{botherref}
\oauthor{\bsnm{Gerschgorin}, \binits{S.}}:
Über die {Abgrenzung} der {Eigenwerte} einer {Matrix}.
Bulletin de l'Académie des Sciences de l'URSS
(6),
749--754
(1931)
\end{botherref}
\endbibitem

\end{thebibliography}

\newcommand{\BMCxmlcomment}[1]{}

\BMCxmlcomment{

<refgrp>

<bibl id="B1">
  <title><p>A {Survey} of {Dynamic} {Spectrum} {Access}</p></title>
  <aug>
    <au><snm>Zhao</snm><fnm>Q</fnm></au>
    <au><snm>Sadler</snm><fnm>B.M.</fnm></au>
  </aug>
  <source>IEEE Signal Processing Magazine</source>
  <pubdate>2007</pubdate>
  <volume>24</volume>
  <issue>3</issue>
  <fpage>79</fpage>
  <lpage>-89</lpage>
</bibl>

<bibl id="B2">
  <title><p>{IEEE} {Standard} for {Information}
  {Technology}–{Telecommunications} and information exchange between systems;
  {Wireless} {Regional} {Area} {Networks} ({WRAN})–{Specific} requirements
  {Part} 22: {Cognitive} {Wireless} {RAN} {Medium} {Access} {Control} ({MAC})
  and {Physical} {Layer} ({PHY}) {Specifications}: {Policies} and {Procedures}
  for {Operation} in the {TV} {Bands}</p></title>
  <aug>
    <au><cnm>IEEE</cnm></au>
  </aug>
  <source>IEEE Std 802.22-2011</source>
  <pubdate>2011</pubdate>
</bibl>

<bibl id="B3">
  <title><p>Wireless {Communications}</p></title>
  <aug>
    <au><snm>Goldsmith</snm><fnm>A</fnm></au>
  </aug>
  <publisher>Cambridge: Cambridge University Press</publisher>
  <pubdate>2005</pubdate>
</bibl>

<bibl id="B4">
  <title><p>Spectrum {Sensing} for {Cognitive} {Radio} : {State}-of-the-{Art}
  and {Recent} {Advances}</p></title>
  <aug>
    <au><snm>Axell</snm><fnm>E.</fnm></au>
    <au><snm>Leus</snm><fnm>G.</fnm></au>
    <au><snm>Larsson</snm><fnm>E.G.</fnm></au>
    <au><snm>Poor</snm><fnm>H.V.</fnm></au>
  </aug>
  <source>IEEE Signal Processing Magazine</source>
  <pubdate>2012</pubdate>
  <volume>29</volume>
  <issue>3</issue>
  <fpage>101</fpage>
  <lpage>-116</lpage>
</bibl>

<bibl id="B5">
  <title><p>A {Survey} of {Spectrum} {Sensing} {Algorithms} for {Cognitive}
  {Radio} {Applications}</p></title>
  <aug>
    <au><snm>Yücek</snm><fnm>T.</fnm></au>
    <au><snm>Arslan</snm><fnm>H.</fnm></au>
  </aug>
  <source>IEEE Communications Surveys Tutorials</source>
  <pubdate>2009</pubdate>
  <volume>11</volume>
  <issue>1</issue>
  <fpage>116</fpage>
  <lpage>-130</lpage>
</bibl>

<bibl id="B6">
  <title><p>A {Review} on {Spectrum} {Sensing} for {Cognitive} {Radio}:
  {Challenges} and {Solutions}</p></title>
  <aug>
    <au><snm>Zeng</snm><fnm>Y</fnm></au>
    <au><snm>Liang</snm><fnm>YC</fnm></au>
    <au><snm>Hoang</snm><fnm>AT</fnm></au>
    <au><snm>Zhang</snm><fnm>R</fnm></au>
  </aug>
  <source>EURASIP Journal on Advances in Signal Processing</source>
  <pubdate>2010</pubdate>
  <volume>2010</volume>
</bibl>

<bibl id="B7">
  <title><p>Robust statistics</p></title>
  <aug>
    <au><snm>Huber</snm><fnm>PJ</fnm></au>
  </aug>
  <publisher>Berlin: Springer</publisher>
  <pubdate>2011</pubdate>
</bibl>

<bibl id="B8">
  <title><p>Radiometric detection of spread-spectrum signals in noise of
  uncertain power</p></title>
  <aug>
    <au><snm>Sonnenschein</snm><fnm>A</fnm></au>
    <au><snm>Fishman</snm><fnm>P.M.</fnm></au>
  </aug>
  <source>IEEE Transactions on Aerospace and Electronic Systems</source>
  <pubdate>1992</pubdate>
  <volume>28</volume>
  <issue>3</issue>
  <fpage>654</fpage>
  <lpage>-660</lpage>
</bibl>

<bibl id="B9">
  <title><p>{SNR} {Walls} for {Signal} {Detection}</p></title>
  <aug>
    <au><snm>Tandra</snm><fnm>R.</fnm></au>
    <au><snm>Sahai</snm><fnm>A.</fnm></au>
  </aug>
  <source>IEEE Journal of Selected Topics in Signal Processing</source>
  <pubdate>2008</pubdate>
  <volume>2</volume>
  <issue>1</issue>
  <fpage>4</fpage>
  <lpage>-17</lpage>
</bibl>

<bibl id="B10">
  <title><p>{SNR} {Walls} for {Feature} {Detectors}</p></title>
  <aug>
    <au><snm>Tandra</snm><fnm>R.</fnm></au>
    <au><snm>Sahai</snm><fnm>A</fnm></au>
  </aug>
  <source>{IEEE} {International} {Symposium} on {New} {Frontiers} in {Dynamic}
  {Spectrum} {Access} {Networks} ({DySPAN})</source>
  <pubdate>2007</pubdate>
  <fpage>559</fpage>
  <lpage>-570</lpage>
</bibl>

<bibl id="B11">
  <title><p>Eigenvalue-based spectrum sensing algorithms for cognitive
  radio</p></title>
  <aug>
    <au><snm>Zeng</snm><fnm>Y</fnm></au>
    <au><snm>Liang</snm><fnm>YC</fnm></au>
  </aug>
  <source>IEEE Transactions on Communications</source>
  <pubdate>2009</pubdate>
  <volume>57</volume>
  <issue>6</issue>
  <fpage>1784</fpage>
  <lpage>-1793</lpage>
</bibl>

<bibl id="B12">
  <title><p>Probability, random variables, and stochastic processes</p></title>
  <aug>
    <au><snm>Papoulis</snm><fnm>A</fnm></au>
    <au><snm>Pillai</snm><fnm>SU</fnm></au>
  </aug>
  <publisher>New York City: McGraw-Hill</publisher>
  <edition>4</edition>
  <pubdate>2002</pubdate>
</bibl>

<bibl id="B13">
  <title><p>Worst-{Case} {Additive} {Noise} in {Wireless}
  {Networks}</p></title>
  <aug>
    <au><snm>Shomorony</snm><fnm>I.</fnm></au>
    <au><snm>Avestimehr</snm><fnm>A.S.</fnm></au>
  </aug>
  <source>IEEE Transactions on Information Theory</source>
  <pubdate>2013</pubdate>
  <volume>59</volume>
  <issue>6</issue>
  <fpage>3833</fpage>
  <lpage>-3847</lpage>
</bibl>

<bibl id="B14">
  <title><p>Recommendation {P}.372-12 : {Radio} noise</p></title>
  <aug>
    <au><cnm>{International Telecommunication Union}</cnm></au>
  </aug>
  <pubdate>2015</pubdate>
  <url>https://www.itu.int/rec/R-REC-P.372-12-201507-I/en</url>
</bibl>

<bibl id="B15">
  <title><p>Low noise electronic system design</p></title>
  <aug>
    <au><snm>Motchenbacher</snm><fnm>CD</fnm></au>
    <au><snm>Connelly</snm><fnm>JA</fnm></au>
  </aug>
  <publisher>Hoboken: Wiley</publisher>
  <pubdate>1993</pubdate>
</bibl>

<bibl id="B16">
  <title><p>Man-{Made} {Noise} {Measurement} {Programme} ({AY}4119) - {Final}
  {Report}</p></title>
  <aug>
    <au><snm>Wagstaff</snm><fnm>A</fnm></au>
    <au><snm>Merricks</snm><fnm>N</fnm></au>
  </aug>
  <pubdate>2003</pubdate>
  <issue>Issue 2</issue>
</bibl>

<bibl id="B17">
  <title><p>Toeplitz and circulant matrices: {A} review</p></title>
  <aug>
    <au><snm>Gray</snm><fnm>RM</fnm></au>
  </aug>
  <publisher>Boston, Delft: Now Publishers</publisher>
  <pubdate>2006</pubdate>
</bibl>

<bibl id="B18">
  <title><p>Über die {Abgrenzung} der {Eigenwerte} einer {Matrix}</p></title>
  <aug>
    <au><snm>Gerschgorin</snm><fnm>S</fnm></au>
  </aug>
  <source>Bulletin de l'Académie des Sciences de l'URSS</source>
  <pubdate>1931</pubdate>
  <issue>6</issue>
  <fpage>749</fpage>
  <lpage>-754</lpage>
</bibl>

</refgrp>
} 





\begin{figure}[h!]
	\includegraphics{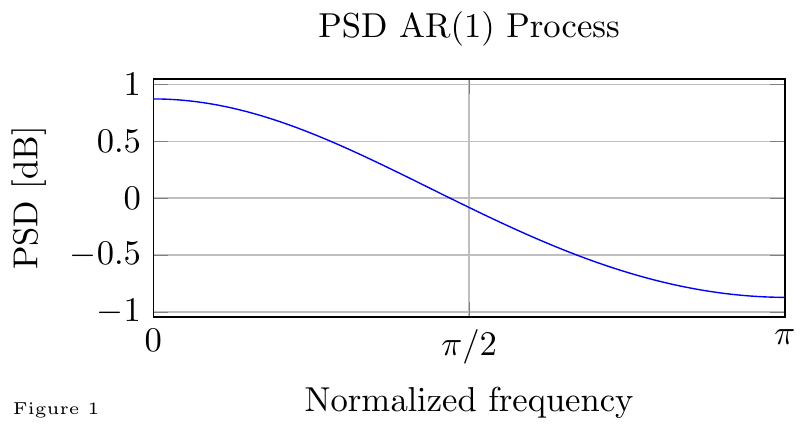}
	\caption{PSD of AR(1) process, which follows  $\eta[n]=0.1\eta[n-1]+ \epsilon [n]$. The frequency is normalized to $f_\text{samples} = 2\pi$.}
	\label{fig:psd_ar_1}
\end{figure}

\begin{figure}[h!]
	\includegraphics{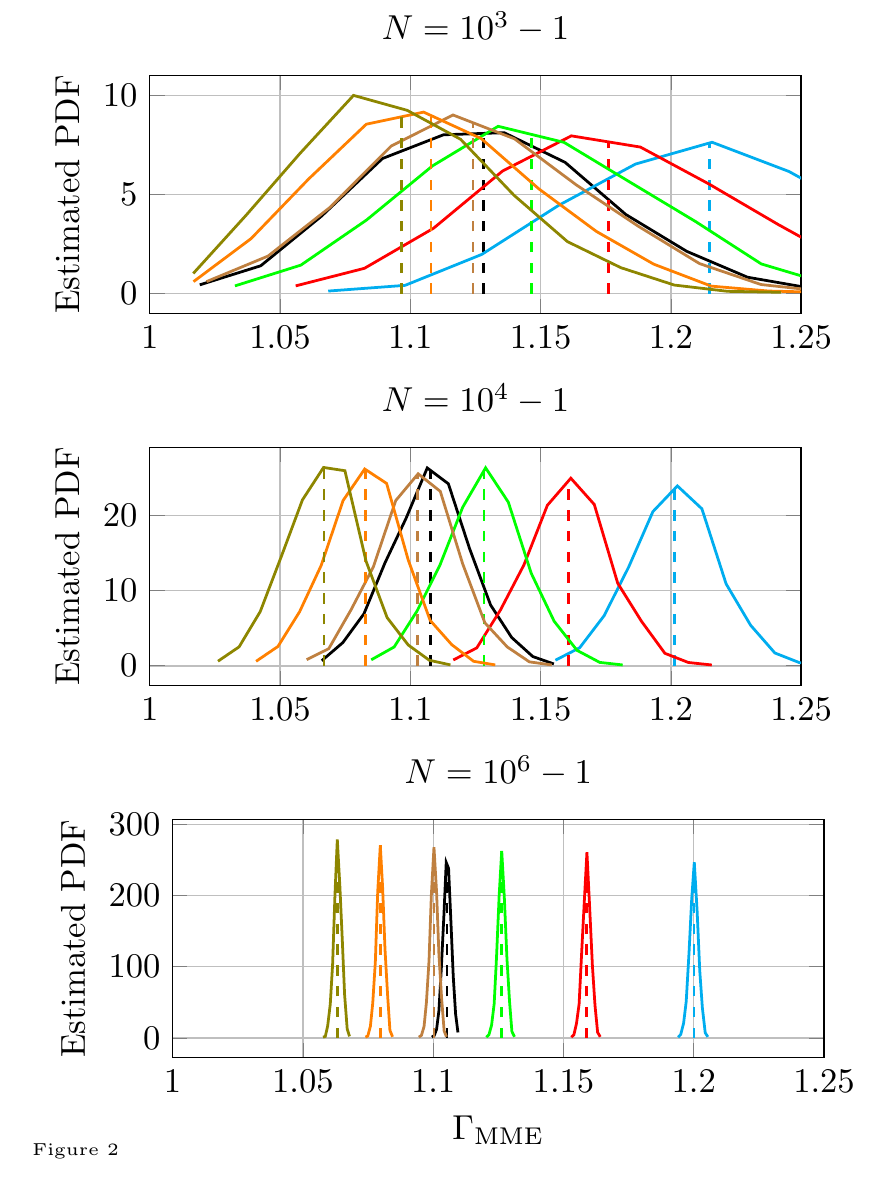}
	\caption {Estimated probability density function for receiver correlation ($p=2$, $Q=0$) and different $N$. Noise correlation factor $\varrho_{\text {max}}^\eta=0.05$. Solid lines represent the histograms, while dashed lines depict the means. The $\hypzero$ case result is shown in black, while the $\hypone$ case results are colored, where the \gls{snr} from right to left is given by $\{-10, -11, \dotsc, -15\}$dB.}%
	\label{fig:pdf_plot_coloured_noise_rec_cor}
\end{figure}

\begin{figure}[h!]
	\includegraphics{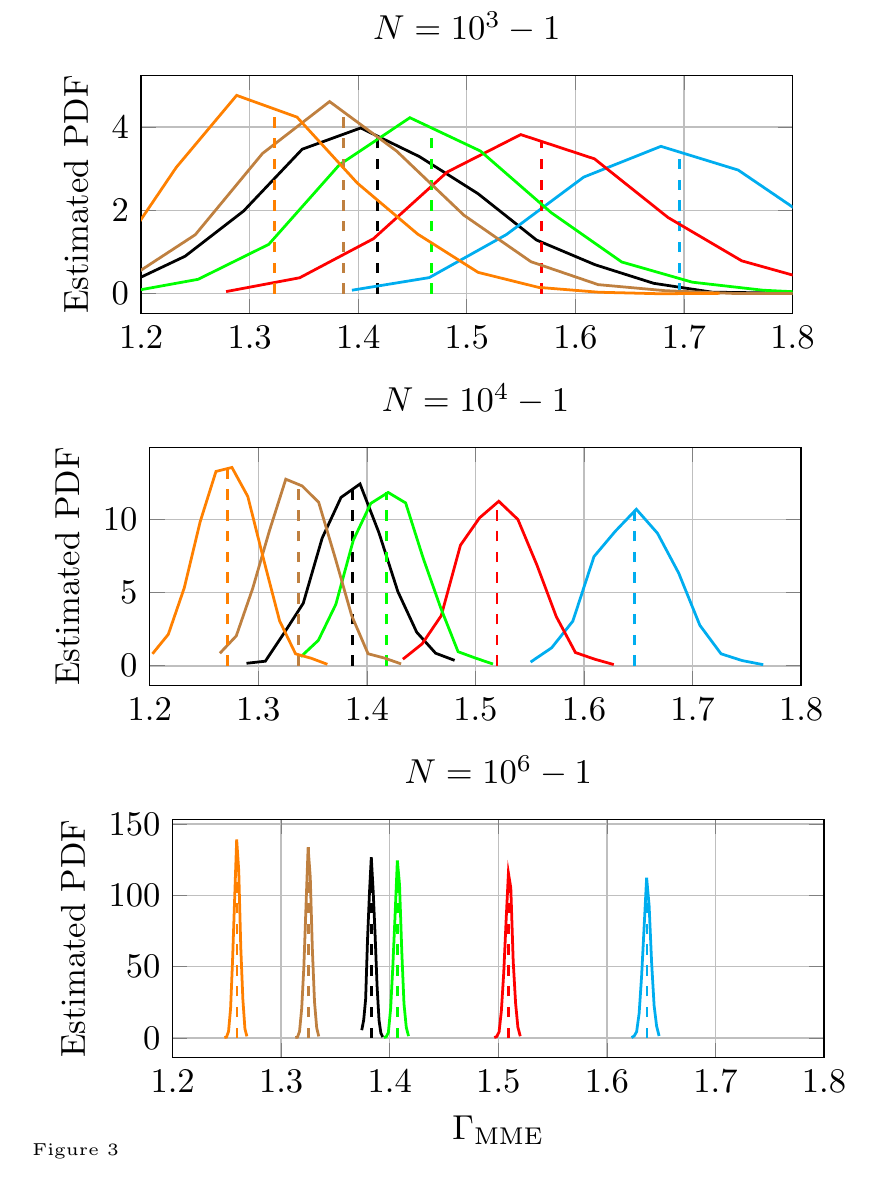}
	\caption {Estimated probability density function for time correlation ($p=1$, $Q=4$) and different $N$. Solid lines represent the histograms, while dashed lines depict the means. The $\hypzero$ case result is shown in black, while the $\hypone$ case results are colored, where the \gls{snr} from right to left is given by $\{-6, -7, \dotsc, -10\}$dB.}
	\label{fig:pdf_plot_coloured_noise_time_cor}
\end{figure}



\begin{table}[h!]
	\caption{System Parameters
	}
	\label{tab:colnoise_sim_par}
	\centering
	{\small
		\begin{tabular}{l c c}
			\hline 
			Parameter & Symbol & Value(s) \\
			\hline 
			Modulation type &  & BPSK \\
			Oversampling factor& $M$&$4$\\
			Number of samples&$N$&$\{999, 9999, 999999\}$\\
			\# of Monte Carlo instances & & 2000\\
			\# of histogram bins&&$12 $\\			
			\hline
		\end{tabular}
	}
\end{table}

\end{backmatter}
\end{document}